\documentclass[%
 reprint,
 amsmath,amssymb,
 aps,
showkeys
]{revtex4-1}

\usepackage{graphicx}% Include figure files
\usepackage{dcolumn}% Align table columns on decimal point
\usepackage{bm}% bold math
\begin{document}

%\preprint{APS/123-QED}

\title{Controlling breath figure patterns on PDMS by concentration variation of ethanol-methanol binary vapors}% Force line breaks with \\

\author{K Nilavarasi$^{a}$}
\author{V Madhurima$^{b}$}%
\affiliation{Department of Physics\\
School of Basic and Applied Sciences\\
Central University of Tamil Nadu\\
Thiruvarur - 610005, Tamil Nadu, India.\\
$^{a}$nilavarasi@students.cutn.ac.in; $^{b}$madhurima@cutn.ac.in
  }%


%\affiliation{
%\affiliation{
 %Third institution, the second for Charlie Author
%\affiliation{%


\date{\today}% It is always \today, today,

\begin{abstract}
In this paper, the self-assembly of condensed droplets on smooth and constrained surfaces under saturated vapor atmosphere of ethanol and methanol binary system is reported. Hexagonally ordered array of pores are obtained on smooth surfaces with saturated vapors of binary liquids without the assistance of any additives. The results show that the addition of small amount of ethanol to methanol plays a role very similar to that of surface active agents in inducing the formation of regular droplet array. The effect of constraints on self-assembled droplet pattern such as movement of contact line and depinning of contact line is also investigated. It is observed that the pore size, pore shape, pore depth and ring diameter are influenced by the atmosphere of binary vapors in addition to the commonly held attribution to the surface tension of the solvent.  Contact angle studies of the patterned substrates showed hydrophobicity with very high adhesiveness to water and Wenzel's state of wetting. 
%\begin{description}
%\verb+\pacs{68.08.?p} 
%\end{description}
\end{abstract}

%Use showkeys class option if keyword
\maketitle

%\tableofcontents

%
\section{Introduction}
\label{intro}
Formation of self-assembled droplet array on substrates (breath figure technique), an efficient bottom-up approach of templating, uses liquid droplets as sacrificial templates for the preparation of micro-structured porous film \cite{du,gong,hernandez,bunz}. This technique involves self-assembly of condensing water droplets into hexagonal arrays on surfaces due to capillary and Marangoni forces. This process typically involves four steps as follows. Step 1. Casting of polymer solution on the substrate surface in the presence of humid air. Step 2. Condensation of water vapor into micro droplets due to evaporative cooling of the solvent. Step 3. Growth and self-assembly of droplets into hexagonal array due to capillary and Marangoni forces. Step 4. Finally, complete evaporation of the solvent and water leads to ordered array of pores \cite{gong,ding,caikang,yang}. The resultant droplet patterns on the polymer films is significantly affected by subtle changes in the casting conditions such as polymer and its structure, solvent, air flow, humidity, surface temperature and substrate \cite{ding}.  

The influence of polymer\cite{Yawen,escale,widawski,yabu}, solvent\cite{escale,bolognesi,yan,fukira,servoli,xu}, temperature\cite{escale,gu}, substrates\cite{escale}, and humidity\cite{escale,stenzel,zou} has been investigated in detail.  However very few reports  investigated the influence of vapor atmosphere on the self-assembly of droplets on surfaces.  For instance, Ding et al., studied the influence of ethanol and methanol vapors on the formation of self-assembled droplet patterns on the polystyrene-block-poly-dimethyl siloxane(PS-b-PDMS) surface. They observed that the surface tension and the enthalpy of solvents in the atmosphere are responsible for the formation of porous films\cite{ding}.  Xiong et.al., reported the formation of microspheres on poly(styrene)-b-poly(butadiene) (PS-b-PB) co polymers using solvent vapor atmosphere.  They also stated that surface tension is a key parameter in the formation of patterns \cite{xiong}.  Bai et al., fabricated polymeric nano/microstructures on poly(9,9-dihexylfluorene) in mixed water and methanol vapor atmospheres \cite{bai}. Zhang et al., reported a modified process for preparing porous films in methanol vapor with conventional polymers, by adding a small amount of surface active agent into the casting solution, such as siloxane- and fluorine-containing block copolymers \cite{aijuan}. The previous studies are performed for either pure methanol or ethanol. It is observed from the above literature that the hexagonal array of pores with methanol vapors are obtained only with the addition of either high surface tension water in the environment or surface active agents to the casting solution. 

Ethanol-methanol binary liquids are interesting because of their complex anomalous behavior that arises due to the molecular structuring of the moieties around each other in solution \cite{madhu}. The inter-facial structuring and hence the complex properties of the binary liquids are significant in affecting the formation of self-assembled droplet formation. In this paper, we present a strategy of obtaining ordered droplet arrays with a binary vapor of ethanol-methanol and a comprehensive description of the role of ethanol in stabilizing methanol droplets during the formation of self-assembled droplet arrays. The influence of constraints of the underlying substrate on the self-assembly process is also investigated.

\begin{figure}[h]
%\resizebox{0.75\textwidth}{!}{%
  \includegraphics{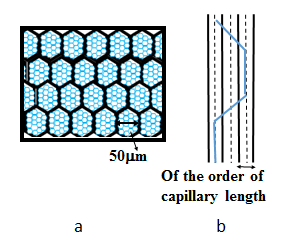}
	\vspace{0.5cm}
%}
\caption{Cartoon shows the effect of constraints with respect to the scale. a. there is no influence of the constraint when the constraint is of the order of $50\mu m$ b. Movement of contact line as shown in blue line when the constraint is of the order of capillary length or less.}
\label{1}
\end{figure}

The presence of constraints on the underlying substrate also influences the resultant droplet patterns.  It is observed from the literature that if the scale of constraints is of the order of $\> 50 \mu m$, then the constraints have no influence on the self-assembly process\cite{escale,park}.  If the scale is of the order of capillary length or less, then the constraints makes the difference in the pattern of droplets self-assembly (Figure~\ref{1}).  This is due to the effect of forces exerted on the contact line of the condensing liquid droplet\cite{feng}. Understanding the dynamics of contact line of condensing liquid droplets is of critically important in the surface science and in many biomedical \cite{carey,boreyko,adam}and industrial applications \cite{adam}. 

On an ideal smooth surface, the contact line remains pinned which is not true in reality \cite{pedro}.  On a real surface, pinning and depinning of contact line occurs due to the liquid-liquid interaction force (L-L force) and solid-liquid interaction force (S-L force). L-L force pulls the contact line towards the interior of the droplet and the S-L force will make the contact line to get pinned on the substrate \cite{berry,marchand,feng}.  This pinning and depinning transition also leads to different evaporation modes viz., constant contact angle mode and constant contact radius mode \cite{feng,wu,kooji,mullar}. The origin of the movement of contact line is still not well understood \cite{feng}.\\

The present work experimentally investigates the influence of saturated vapors of ethanol-methanol binary vapors (over entire concentration regions) on the self-assembled droplet patterns on PDMS surfaces over both smooth and constrained substrates. An attempt is also made to study the influence of constraints on the underlying surface and tried to explain the self-assembly of droplets in terms of intermolecular interactions of the binary liquid vapors.
\section{Experimental details}
\label{sec:2}
\subsection{Materials}
\label{subsec:2.1}
Polydimethyl siloxane (PDMS)  from SIGMA Aldrich and chloroform (purity $99.9 \% $) purchased from EMPLURA are used as polymer and solvent respectively. Methanol and ethanol are purchased from Sigma Aldrich and Merck Emplura respectively. The purities of methanol and ethanol are $99\%$.  Commercially available SONY recordable DVDs are used as substrates. The DVD disc is cut into pieces of $1 cm \times 1 cm$. The two polymer layers are carefully peeled off.  The polycarbonate support and the smooth polymer layer (the other one) are used as constrained and smooth surface respectively.  
\subsection{Preparation and characterization of binary liquids}
\label{subsec:2.2}
Binary mixtures of ethanol and methanol are prepared by mixing appropriate volume of each liquid in an airtight stopper glass bottle. The accuracy of the binary system is $\pm 0.1 mg$ and they are measured using a weighing balance. Surface tension of the binary systems for all concentration range are measured using Rame-Hart contact angle goniometer.  Pendant drop method is used in calculating the polar and dispersive parts of surface tension.  \\ 

The infra-red spectra of the binary system in the spectral range from $400$ to $4000 cm^{-1}$ are measured by FTIR Spectrometer of Perkin Elmer.  The dielectric studies are carried out using 40 GHz Vector Network Analyzer with Dielectric Assessment Kit of Rhode and Scharwz. The refractive index measurements of the binary system are carried out using Abbe's refractometer.  Calibration is performed by measuring the refractive indices of doubly distilled water using Na light. All the experiments are done at room temperature. \\

\subsection{Self-assembly of binary liquid droplets}
\label{subsec:2.3}
\begin{figure}[h]
%\centering
\resizebox{0.5\textwidth}{!}{%
  \includegraphics{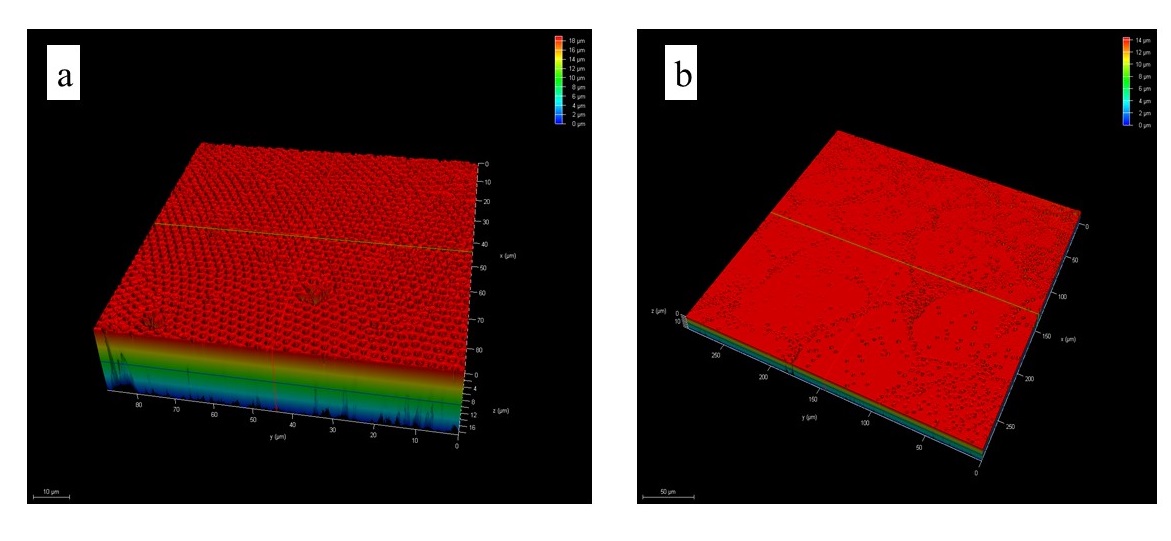}
	\vspace{0.5cm}
}
\caption{3D CLSM images of chloroform treated a. smooth and b. constrained substrates patterned under water-vapor.  }
\label{1-b}
\end{figure}

In the current experiments, mixture of ethanol and methanol at various volume ratios are chosen to produce vapor atmospheres. $3 ml$ of mixture of binary liquids is added in a petri-dish and kept inside a sealed glass chamber to form a saturated vapor atmosphere. The smooth and the constrained surfaces are positioned at least $1cm$ above the liquid. About $300$ micro liter of PDMS-chloroform solution is casted on the surface of the substrates using a micro-syringe under the binary liquid saturated vapor atmosphere at $ ~ 26^{0}$C.  After the complete evaporation of the solvent, a thin layer of polymer is left on the substrate surface. \\ 
\begin{figure}[h]
%\resizebox{0.75\textwidth}{!}{%
  \includegraphics{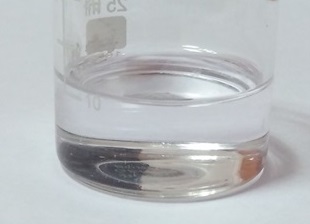}
	\vspace{0.5cm}
%}
\caption{Photograph of DVD dipped ethanol solution after 30 minutes which shows no precipitates}
\label{1-a}
\end{figure}
To check the dissolutivity of chloroform with DVD, two blank experiments are performed. The substrates are treated with ethanol and checked for precipitates.  It is found that with the given time of 1 minute, there is no formation of precipitate.  To further confirm the dissolutivity, the DVD discs are treated with chloroform and the treated discs are used for patterning under saturated vapors of water.  It is observed that the patterns formed on the chloroform treated smooth and constrained surfaces is similar to untreated patterned surfaces. Therefore it is confirmed that within the time span of the experiment, the chloroform doesn't dissolve the upper layer of DVD discs.  The supporting images are shown in (Figure~\ref{1-a} and ~\ref{1-b}) \\

\subsection{Characterization of patterned surfaces}
\label{subsec:2.4}
The patterns formed on the substrate surface are characterized using the Leica Confocal Scanning Electron Microscope TCS SP8 model. The transmission mode with the excitation wavelength of 561 nm is used to capture the images of formed patterns. Rame Hart contact angle goniometer equipped with a charge coupled device camera along with an advanced drop image program is used to measure the contact angle of water with the surfaces. $3\mu l$ drop of distilled water is placed on the substrates through a microsyringe which is placed onto the stage. The measurement is recorded after the drop reaches equilibrium.  The substrate is moved to allow another drop to be placed on the substrate. Atleast 10 measurements are taken and all are reproducible up to $\pm 2^{0}$. 

\section{Results and discussion}
\label{sec:3}
Precise experimental conditions are required for self-assembly of droplet patterns since evaporation and film formation occur in few seconds.  To avoid the possibility of uncertainties with dynamic method of self-assembly of droplets, the static method explained in the experimental section is employed to fabricate the structures on smooth and constrained substrates. This method is quite robust under saturated vapor of water.  Few works have systematically examined the self-assembly of droplet patterns on smooth surfaces in ethanol and methanol environment \cite{ding}.  In the present case the binary system of ethanol and methanol are employed as vapor atmosphere during the process of self-assembly of droplets. \\

\begin{table*}
\small
\caption{Variation of properties of ethanol-methanol binary system with the concentration of methanol}
\label{tab:1}
%\centering
\begin{tabular*}{\textwidth}{@{\extracolsep{\fill}}llllllllll}
%\begin{tabular}{l c c c c c c c c c}
\hline
Perc. of  & Surface tension                                        &Refractive &Dielectric               &Dielectric \\
methanol  & $(mJ/m^{2})$($\pm 0.02$)                    &index ($\pm 0.001$)      &permitivity ($\pm 2\%$)  &loss($\pm 2\%$)\\ \hline
0         &22.00      											  		         &1.360 		 &24.80
	&8.52
\\
10        &22.98                                           &1.363 		 &24.75
		&10.00
\\
12        &22.14                                           &1.362 		 &24.65
		&6.41
\\
14        &19.30                                           &1.362 		 &24.90
			&8.16
\\
16        &19.46                                           &1.360 		 &25.33
		&9.08
\\
18        &20.70                                           &1.358 		 &25.71
		&9.98
\\
19        &19.88                                           &1.358 		 &25.58
			&3.35
\\
20        &20.70                                           &1.357 		 &25.67
		&4.07
\\
21        &19.01                                           &1.355 		 &26.72
		&4.50
\\ 
22        &20.07                                           &1.355 		 &26.52
		&11.33
\\
24        &19.76                                           &1.355 		 &26.85
			&11.46
\\
26        &19.28                                           &1.353 		 &27.01
			&11.20
\\
28        &19.94                                           &1.352 		 &27.18
		&11.79
\\
30        &21.56                                           &1.352 		 &24.66
		&9.26
\\
40        &21.47                                           &1.347 		 &24.79
			&9.00
\\
50        &21.53                                           &1.342 		 &27.58
	&8.15
\\
60        &21.45                                           &1.337 		 &29.20
	&7.56
\\
70        &21.68                                           &1.333 		 &28.31
		&8.06
\\
80        &21.53                                           &1.331 		 &29.82
		&7.33
\\
90        &21.48                                           &1.324 		 &29.09
			&7.67
\\
100       &22.10                                           &1.327 		 &34.38
		&8.88
\\ \hline
\end{tabular*}
\end{table*}

Most alcohols interact via strong repulsive forces and weak attractive forces. But an important part of the interaction is the strong, directional hydrogen bonds they form which are important in determining its structure. To understand the intermolecular interactions between methanol-ethanol, the properties like surface tension, refractive index and static dielectric permittivity of the binary system over the entire concentration range are studied and tabulated in Table~\ref{1}.  From our recent work on methanol-ethanol binary system, it is observed that all the properties show deviation from ideal behavior at $10\%$ to $30\%$ of methanol. The deviation in the properties of binary system is attributed to the increase of hydrogen bond strength with increase in methanol concentration.  At higher concentration of ethanol, the London dispersion forces dominate over the hydrogen bonds \cite{madhu}. With these deviations, it is also expected a change in self assembly pattern  of droplets over the substrates. \\

\begin{figure}[h]
\resizebox{0.5\textwidth}{!}{%
  \includegraphics{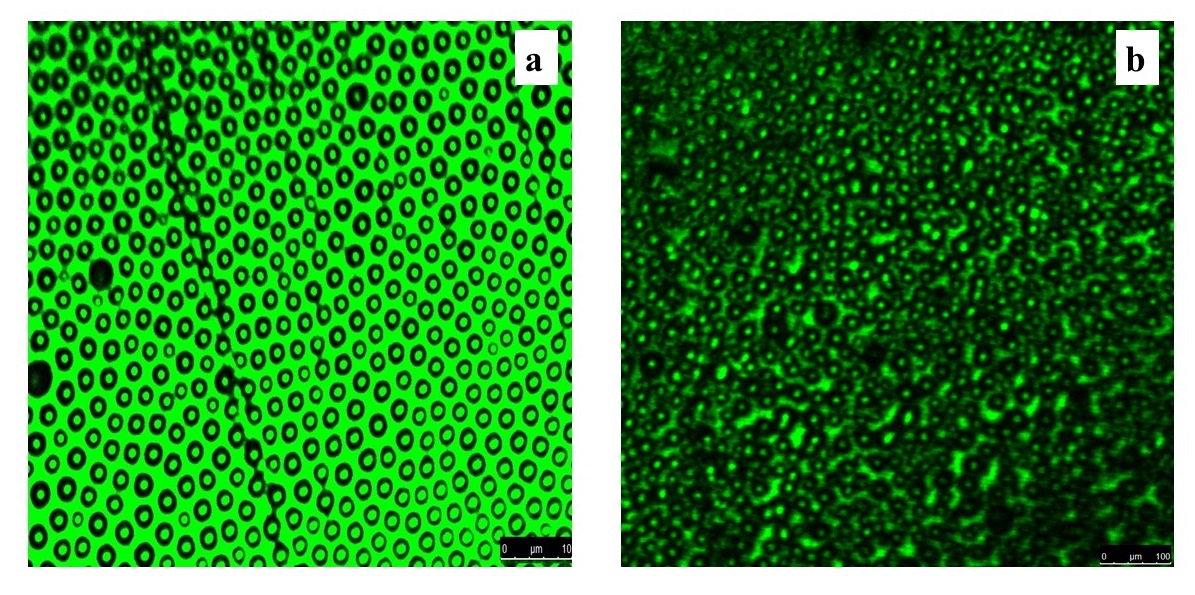}
	\vspace{0.5cm}
}
\caption{CLSM images of PDMS films prepared on smooth substrate under a) pure ethanol and b)Pure methanol .}%
\label{2}
\end{figure}

An attempt at explaining self assembly of liquid droplets in terms of intermolecular interactions and the effect of constraints on the pattern formation, a drop of PDMS-chloroform solution is casted over smooth and constrained substrates in the vapor environment of ethanol-methanol binary system. The solvent is allowed to evaporate. When the solvent evaporates completely, a porous film is left on the substrate. The experiment is repeated over entire concentration range of methanol in ethanol in steps of $10\%$ of volume. \\

\subsection{Pattern formation on smooth and constrained surfaces}
\label{subsec:3.1}
\begin{figure}[h]
\resizebox{0.5\textwidth}{!}{%
  \includegraphics{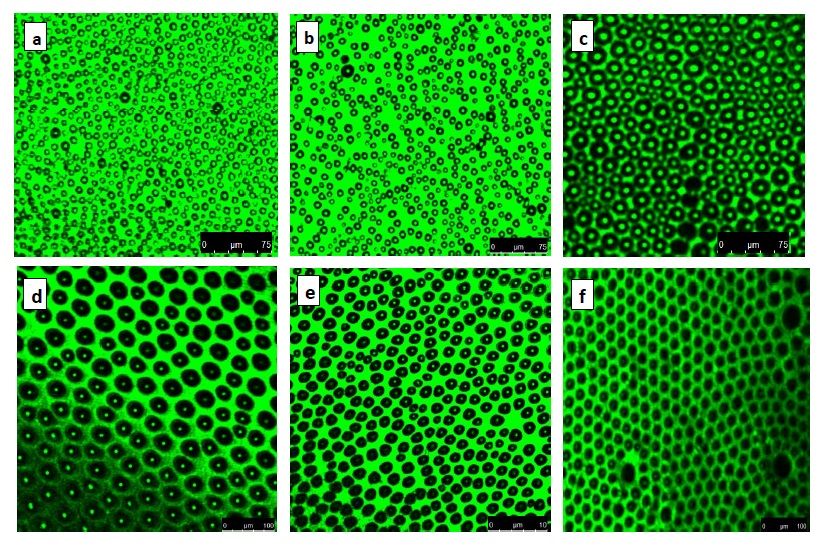}
	\vspace{0.5cm}
}
\caption{CLSM images of PDMS films over smooth substrate prepared in various concentration of binary vapors of ethanol-methanol. a)$10\%$ of methanol  b) $20\%$ of methanol c) $30\%$ of methanol d)$40\%$ of methanol e) $50\%$ of methanol f) $90\%$ of methanol.}%
\label{3}
\end{figure}

\begin{figure}[h]
\resizebox{0.5\textwidth}{!}{%
  \includegraphics{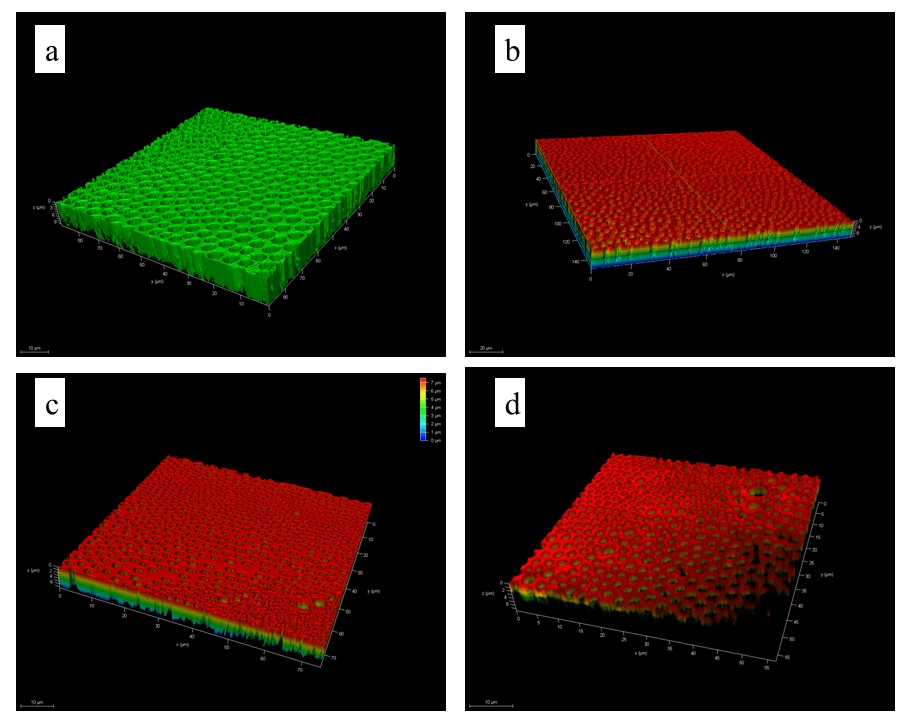}
	\vspace{0.5cm}
}
\caption{3D CLSM images of PDMS films over smooth substrate prepared in various concentration of binary vapors of ethanol-methanol. a)$40\%$ of methanol  b) $50\%$ of methanol c) $70\%$ of methanol d)$90\%$ of methanol.}%
\label{3-a}
\end{figure}

\begin{figure}[h]
\resizebox{0.5\textwidth}{!}{%
  \includegraphics{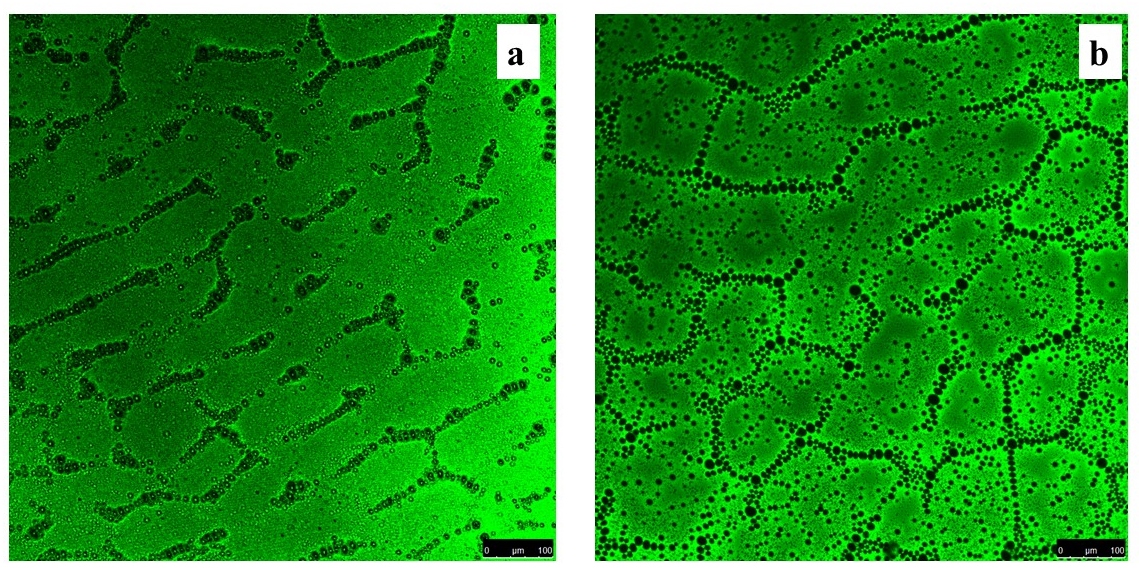}
	\vspace{0.5cm}
}
\caption{CLSM images of PDMS films prepared on constrained substrate under a) pure ethanol and b)Pure methanol.}%
\label{5}
\end{figure}

\begin{figure}[h]
\resizebox{0.5\textwidth}{!}{%
  \includegraphics{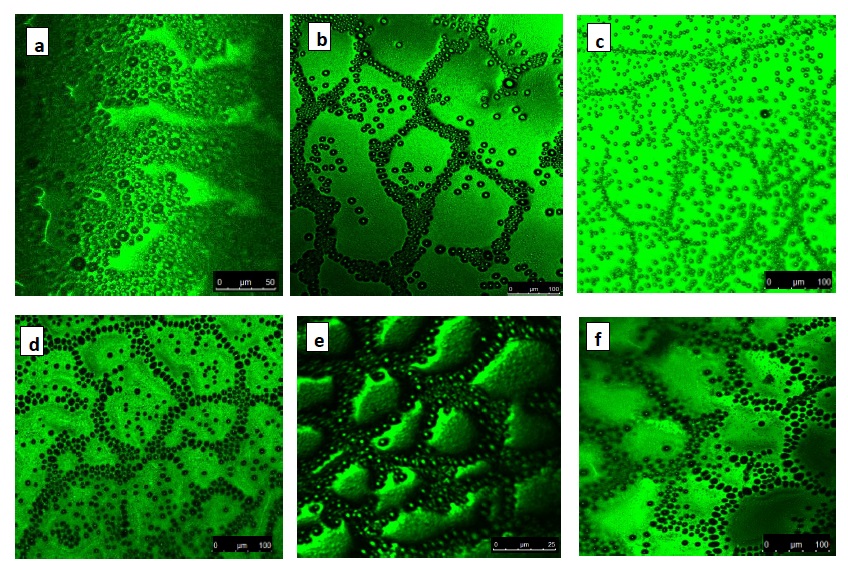}
	\vspace{0.5cm}
}
\caption{CLSM images of PDMS films over constrained substrate prepared in various concentration of binary vapors of ethanol-methanol. a)$10\%$ of methanol  b) $30\%$ of methanol c) $60\%$ of methanol d)$70\%$ of methanol e) $80\%$ of methanol f) $90\%$ of methanol.}%
\label{6}
\end{figure}

\begin{figure}[h]
\resizebox{0.5\textwidth}{!}{%
  \includegraphics{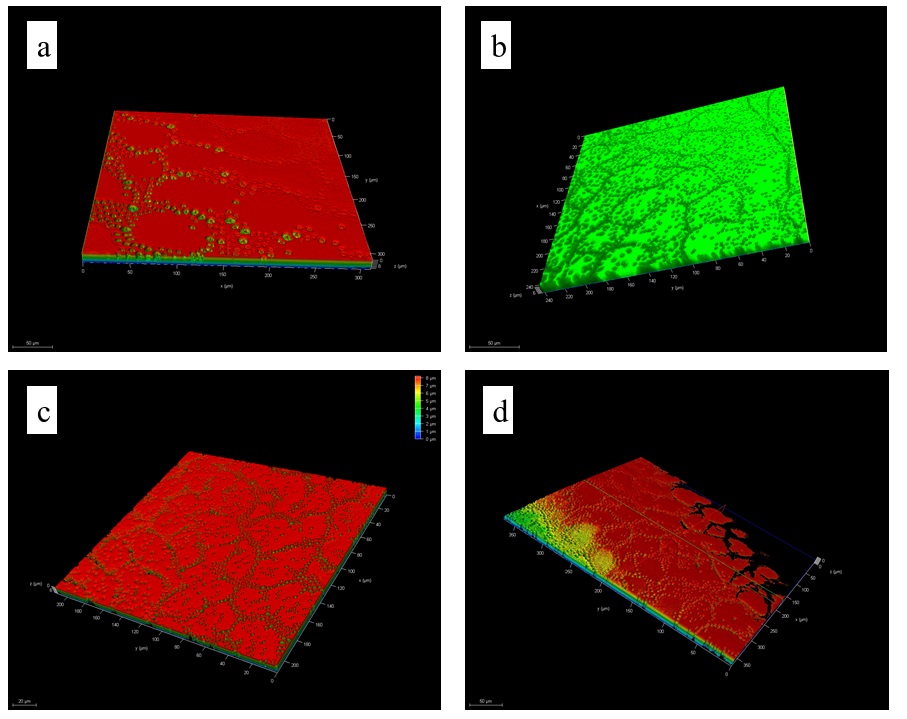}
	\vspace{0.5cm}
}
\caption{3D CLSM images of PDMS films over constrained substrate prepared in various concentration of binary vapors of ethanol-methanol. a)$50\%$ of methanol  b) $60\%$ of methanol c) $70\%$ of methanol d)$90\%$ of methanol.}%
\label{4-a}
\end{figure}

\begin{figure}[h]
\resizebox{0.5\textwidth}{!}{%
  \includegraphics{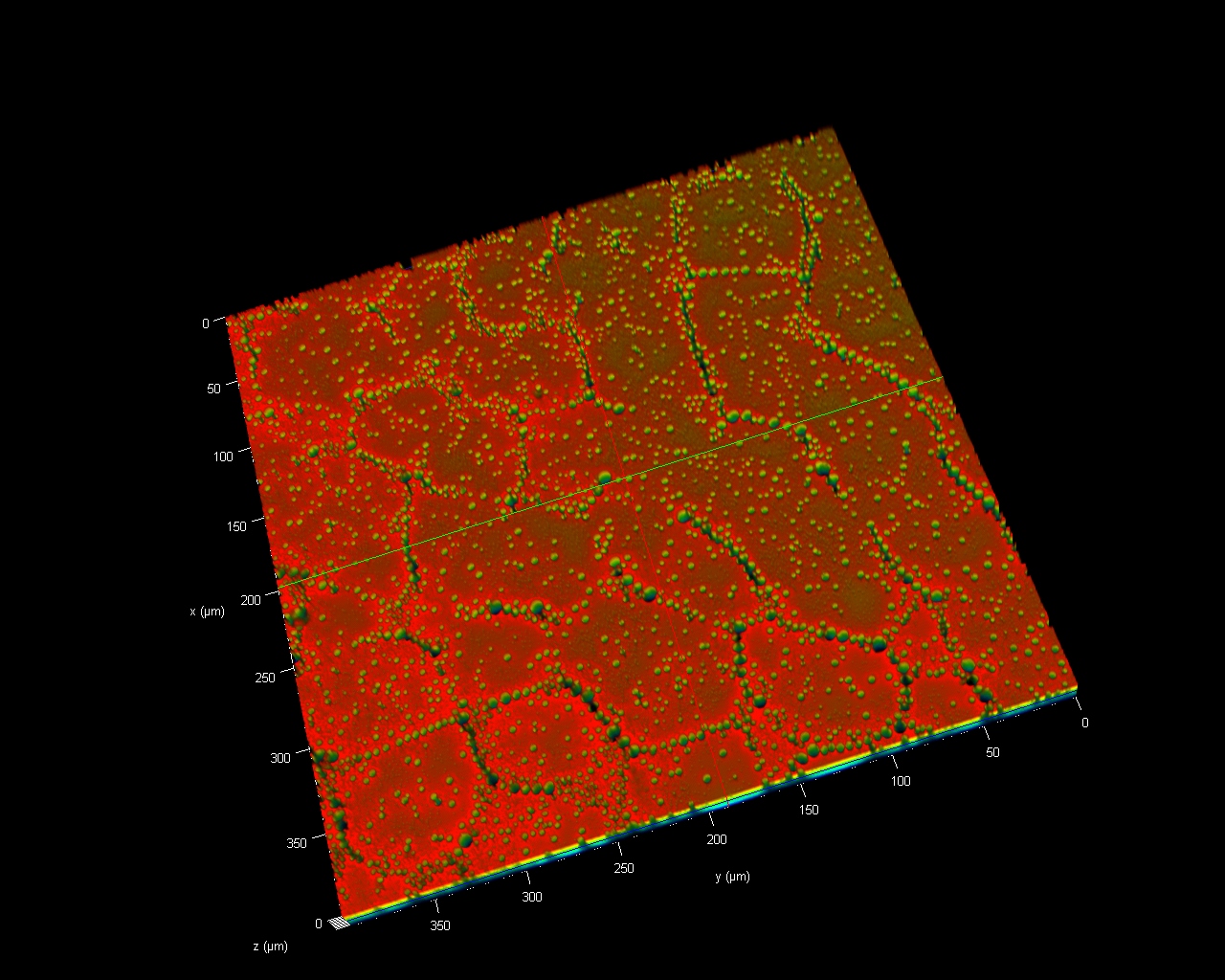}
	\vspace{0.5cm}
}
\caption{3D CLSM image of PDMS films prepared in pure methanol vapor on constrained surface.}%
\label{5-a}
\end{figure}

The Confocal Laser Scanning microscope (CLSM) images of thus patterned smooth and constrained surfaces for all molar fraction of methanol in the binary system are shown in Figures~\ref{2},~\ref{3},~\ref{5},~\ref{6},~\ref{9} and ~\ref{10} and the 3D CLSM images are also shown in Figures~\ref{3-a},~\ref{5-a},~\ref{4-a},~\ref{9-a} and ~\ref{10-a}. The shape, size and overall pattern of the pores prepared under various concentration range of ethanol and methanol are investigated. It is observed that there are variations in the size and shape of the pores on PDMS film patterned under various volume fractions of binary vapors of methanol and ethanol.  \\

Patterns on smooth surface:\\

The images of smooth surface patterned under pure and binary mixtures of ethanol and methanol are shown in Figure~\ref{2} and Figure~\ref{3} respectively. The size of the pores on the PDMS film prepared under various concentration of methanol vary according to the surface tension of the liquids forming saturated vapor.  The shape of the pore openings also vary from polygon to circular. Pores with irregular openings are formed on the polymer film patterned under various concentrations of methanol. 3D CLSM images are also shown in Figure~\ref{3-a} \\

Figure~\ref{2}a shows the pore formation on the smooth surface patterned under vapors of pure ethanol.  It indicates that the ethanol vapors favor the hexagonal pore formation on PDMS films.  The film is densely covered with pores that are nearly $3.25 \mu m$ in size. The pores are separated from each other by a distance of nearly $1 \mu m$.  With the introduction of $10\%$ of methanol, the patterns on the film slightly deviates from their close arrangement of pores.  Figure~\ref{3}a indicates that, the introduction of $10\%$ of methanol decreased the pore size and nearest neighbor pore distance to $1.65 \mu m$ and $800 nm$ respectively. Therefore, it is inferred from the observations that the pore diameter can be controlled by the percentage of methanol in the binary system forming saturated vapor atmosphere. This implies that the pore diameter is a function of surface tension of the binary system. \\

Further increase in the concentration of methanol results in increased pore size and pore density as shown in Figure~\ref{3}c and d.  It is also observed from the figure that at $30\%$ of methanol, the pores are self-assembled into nearly hexagonal arrays.  Around $40\%$ of methanol, the film is observed to have highly ordered hexagonal structures with a pore size of $4.4 \mu m$.  This result indicates that intermolecular interaction between ethanol and methanol molecules increases with increase in methanol concentration.  Above $40\%$ of methanol, the complex behavior of the binary system saturates indicating the stability in their intermolecular forces which is also confirmed from the surface tension measurements as shown in Table~\ref{1}.  \\

Self-assembly of droplet pattern is a surface tension driven phenomenon. Liquids of same surface tension result in similar pattern formation on unconstrained surfaces.  Thus the patterns formed on smooth surface are similar for the concentration range of $40\%$ to $90\%$ of methanol as expected.  A gradual change in the polymeric pore morphology is observed when the saturated vapor atmosphere is composed of pure methanol as shown in Figure~\ref{2}b. \\

From the above results, it is observed that in smooth surface, pure ethanol forms hexagonal array of pores.  However, with the addition of methanol, pore patterns deviates from the hexagonal array.  On the other hand, i.e., with pure methanol highly disordered patterns are observed.  With the addition of small amount of ethanol, the droplets self-assemble themselves into hexagonal array of pores. Thus the addition of small amount of ethanol to methanol plays a role very similar to that of surface active agents in inducing the formation of regular droplet array.  Similar analysis is performed with the constrained surfaces to investigate the role of ethanol-methanol vapors.  Interesting results are observed as follows. \\

Patterns on constrained surface:\\
\begin{figure}[h]
\resizebox{0.5\textwidth}{!}{%
  \includegraphics{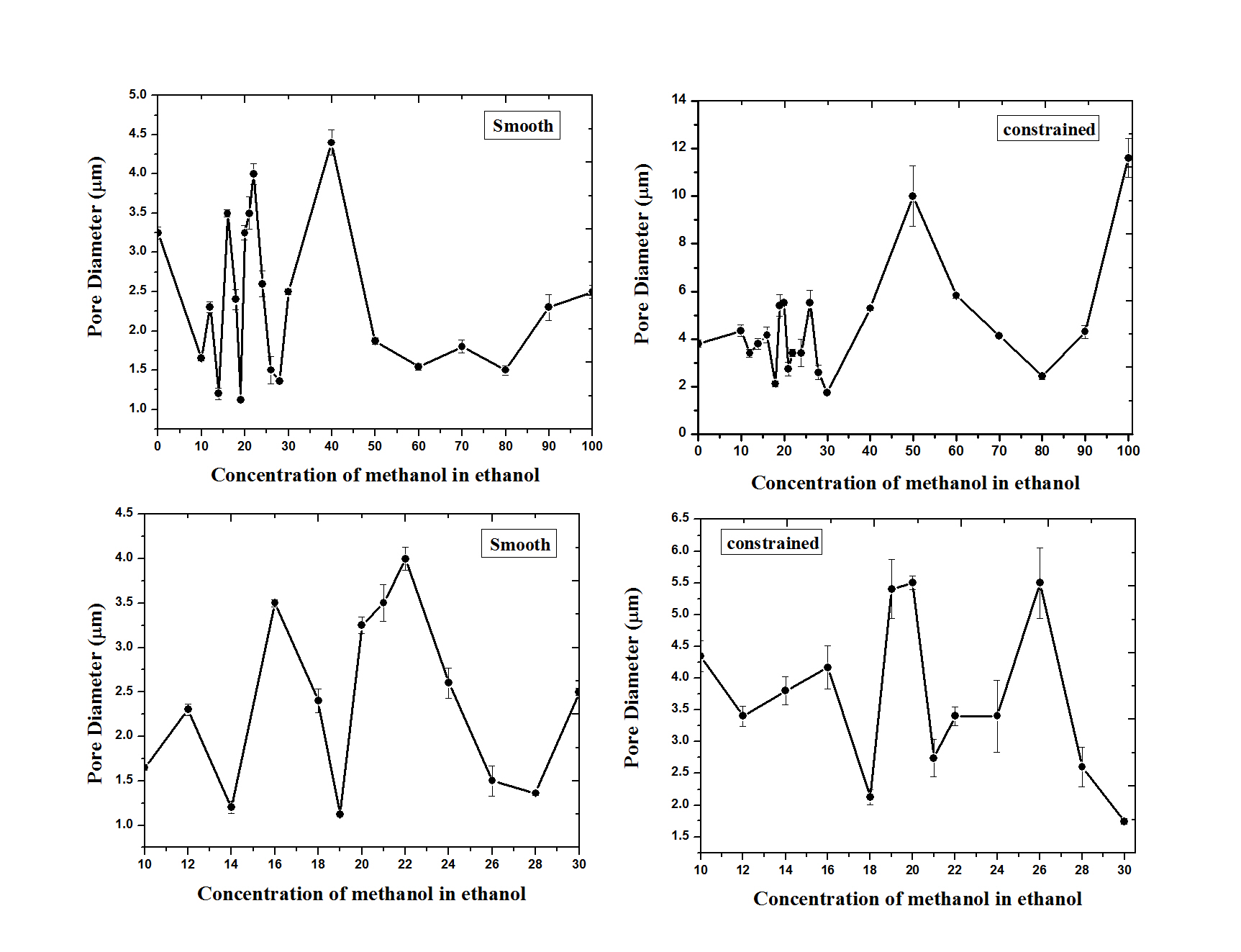}
	\vspace{0.5cm}
}
\caption{Variation of pore diameter for smooth and constrained surfaces patterned with various concentration of methanol in ethanol.(Top: entire concentration region; bottom: concentration region $10\%$ to $30\%$ of methanol).}%
\label{4}
\end{figure}

Figure~\ref{5}a shows the self assembled droplet pattern formed on the constrained surface under saturated vapor of pure ethanol. It is observed that the film is covered with distorted unclosed rings of pores. The ring size and pore size are approximately $36.4 \mu m$ and $3.8 \mu m$.  Pores of size $~0.5 \mu m$ are found inside the distorted rings (size of the pores versus concentration of methanol is shown in Figure~\ref{4}).  On the addition of methanol, the pores self-assembles into patches of hexagons. Further addition of methanol favors the formation of hexagonal rings.  The ring shape attain its hexagonal alignment at higher concentrations of methanol (Figure~\ref{6}). The pore density inside the ring increases with increase in methanol concentration. 3D CLSM images for various concentration of methanol are shown in Figure~\ref{4-a}\\

 With pure methanol as saturated vapor, the rings are found to be highly ordered (3D image of pure methanol on constrained surface is shown in Figure~\ref{5-a}) and densely populated with pores of size $~1\mu m$ (Figure~\ref{5}b) whereas ethanol results in distorted pore array. This results are completely opposite to what is observed in smooth surface.  On constrained surface, despite the role of surface tension of ethanol-methanol binary system, the effect of constraints on the underlying surface dominates in the formation of pore patterns. \\ 

\subsection{Formation mechanism of self-assembled droplet patterns under saturated binary vapors}
\label{subsec:3.2}
To explain the differences in pattern formation prepared under the saturated vapor atmosphere of ethanol-methanol binary system, the physical properties of liquids forming the saturated vapor atmosphere, solvent and polymer are taken into account. The droplet placed on a surface spreads to a certain extent until the equilibrium is reached.  At equilibrium, the total free energy is minimized and the droplet shape stops varying. It is well known that the droplet shape is determined by the inter-facial tension between the polymer and the droplet and the intermolecular interactions within the droplet \cite{yin}.   And the spreading co-efficient in terms of these two factors is given by, 
	\begin{equation}
	S = \gamma_{p}-(\gamma_{d}+\gamma_{pd})
	\label{eq:}
	\end{equation}	
where $\gamma_{p}$, $\gamma_{d}$ and $\gamma_{pd}$ are surface tension of the polymer solution, surface tension of droplet, and inter-facial tension of polymer solution and droplet respectively.  The spreading co-efficient is used to characterize the spreading of droplets and it defines different regimes of wetting.  The equilibrium contact angle, called the Young's contact angle is given by \cite{young},
	\begin{equation}
	\gamma_{d} cos \theta = \gamma_{p}-\gamma_{pd}
	\label{eq:}
	\end{equation}
For a large number of droplets, $\theta $ is the inner angle of the droplet-vapor interface that makes with the polymer layer.  The relation connecting contact angle and spreading co-efficient is obtained by combining eqn.1 and eqn.2 and is given as,
	\begin{equation}
	cos \theta = 1+(S/\gamma_{d})
	\label{eq:}
	\end{equation}\\
The spreading parameter $S>0$ indicates complete wetting and $S<0$ indicates partial wetting. In the present work, the surface tension of polymer solution is kept constant. The surface tension of the droplets is a varying quantity.  With the variation in $\gamma_{d}$, the inter-facial tension of the polymer solution and droplet also changes a little.  Since ethanol and methanol are miscible in chloroform, the inter-facial tension of polymer solution and droplet are assumed to be small.  The inter-facial tension of the polymer solution is measured using the pendant drop method and is measured to be $17.11\times10^{-3} J/m^{2}$.  The surface tension of binary liquid of various concentrations of methanol are shown in Table~\ref{1}  \\

Therefore, the spreading parameter for the droplet of binary liquid for all concentrations can be expected to be positive.  The droplet impinging on the polymer surface will tend to spread.  The formation of intervening polymer layer due to condensation of the saturated vapor prevents the coalescence of thus formed droplets.  As the droplets grow, they are attracted each other through the capillary forces and after complete evaporation of the solvent, the self-assembled droplet arrays are formed. The surface tension of the template droplets determines the shape of the pores. The larger surface tension leads to more spherical pores and deformation from spherical shape occurs with the decrease in surface tension. \\

\begin{figure}[h]
\resizebox{0.5\textwidth}{!}{%
  \includegraphics{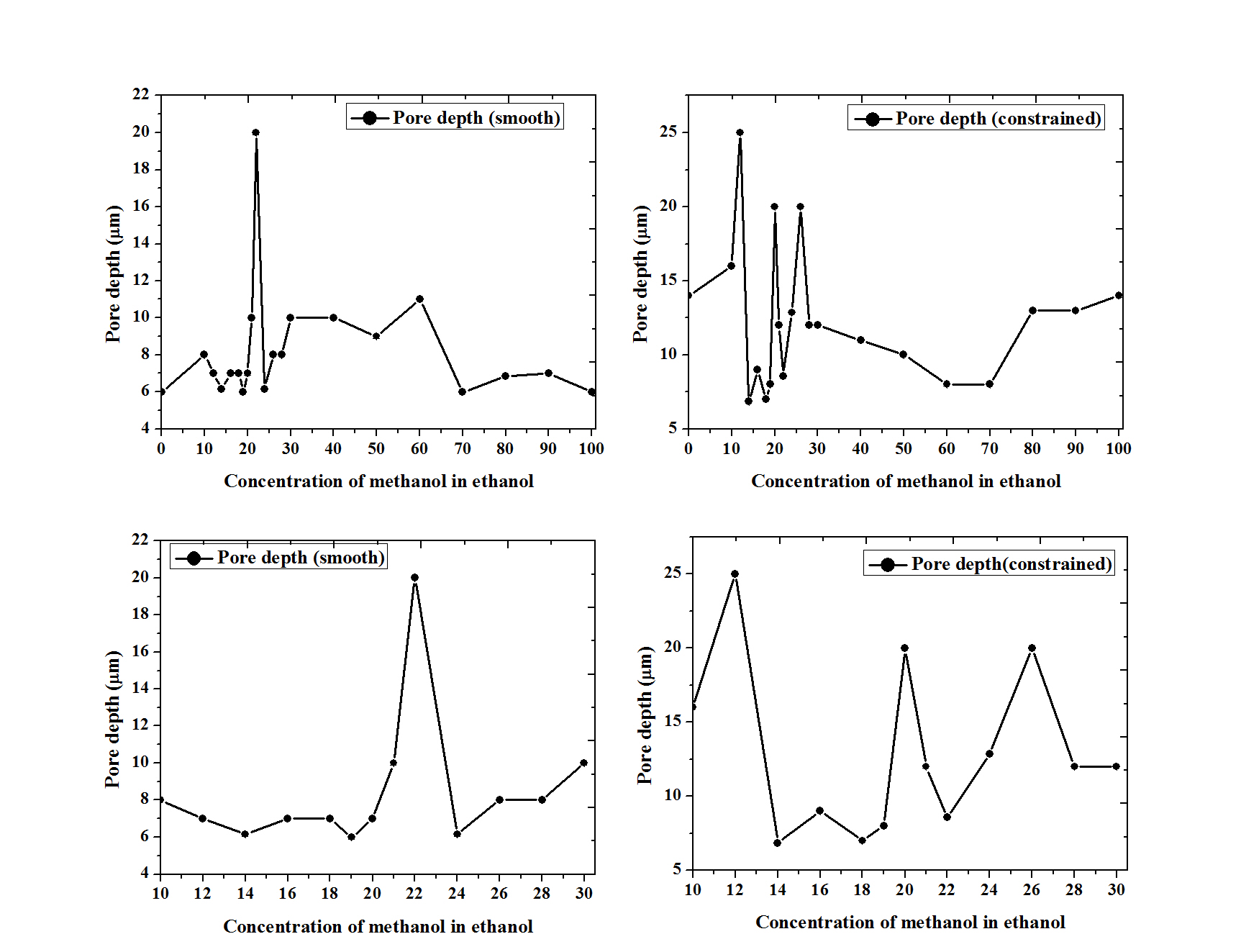}
	\vspace{0.5cm}
}
\caption{Variation of pore depth for smooth and constrained surfaces patterned with various concentration of methanol in ethanol. Uncertainty is found to be less than the symbol size. (Top: entire concentration region; bottom: concentration region $10\%$ to $30\%$ of methanol)}%
\label{7}
\end{figure}
Ethanol and methanol with surface tension $22\times10^{-3} J/m^{2}$ and $22.1\times10^{-3} J/m^{2}$ leads to deformed spherical pores. Figure~\ref{2} shows the formation of deformed pores with methanol droplets and shapes close to sphere with ethanol droplets. With binary liquid droplets, the deformation in spherical shape decreases with increase in methanol concentration and at $40\%$ of methanol, the pores are almost spherical(Figure~\ref{3}). Further increase in methanol concentration leads to increase in deformation of spherical shape. Apart from shape of the pores, the size and depth of the pores also show variations with varying methanol concentration (Figure~\ref{4}, \ref{7}). The variation in size arises from the difference in evaporation enthalpy of the saturated vapor. Evaporation enthalpy determines the amount of vapor to be condensed during the self-assembly of droplets\cite{ding}. Ethanol with its greater evaporation enthalpy leads to smaller pores compared to  the pores formed from methanol vapors. This is due to the fact that lower value of enthalpy leads to condensation of large amount of saturated vapor.  \\

\subsection{Effect of roughness on self-assembly process}
\label{subsec:3.3}
In order to further probe the differences in pore morphology of constrained surfaces, the effect of depinning/pinning of contact line is investigated in detail. A bare/smooth surface uniformly coated with polymer layer has no effect on the movement of contact line of condensed droplets.  Whereas the spreading of liquid on surfaces is strongly influenced by chemical heterogeneities and roughness of the surface \cite{gennes,ramos}.  The contact angle of a drop on a heterogeneous surface is determined solely by the interactions occurring at the three phase contact line. The wetting of heterogeneous surfaces is controlled by the three phase structure at the contact line and not by the inter-facial contact area \cite{extrand}. \\   

The interaction between the fluid interface and the heterogeneities results in the movement of contact lines.  This dynamics of contact line is determined by the upper length scale of the droplets and the lower length scale set by the characteristic size of the constraints/grooves\cite{deniz,jacco}. For a surface with infinitely long grooves close to each other like that of number of lines normal to the axis in a capillary, the contact line can lie either parallel to the grooves or at an angle $\phi$ to the grooves. The contact line gets pinned when it is parallel to the grooves and it moves continuously without any pinning when it lies at an angle $\phi$ \cite{gennes}. \\ 

When the contact line moves over the surface there are often points that remain pinned, inducing the contact line to suddenly jump to a new position \cite{marmur,mi}.  Velocity of the contact line based on calculating the energy dissipation per length of unit line is given by Brochart-Wyatt and de Gennes as \cite{wyart,duursma},
\begin{equation}
V = \frac{\theta\sigma}{6\eta \ln(L/a)}(cos\theta_{o}-cos\theta)
\label{eq:}
\end{equation}
where $\theta$ is the dynamic contact angle, $\sigma$ is the liquid-vapor surface tension , $\theta_{o}$ is the equilibrium contact angle, L/a is the ratio of macroscopic to microscopic length scales and $\eta$ is the dynamic viscosity.  From the above equation it is clear that the contact line velocity is determined by the dynamic viscosity and surface tension of the liquid.  The contact line velocity is high for liquids of high surface tension. So it is expected a difference in pattern formation on constrained surface with condensing droplets of various surface tension. \\
\begin{figure}[h]
\resizebox{0.5\textwidth}{!}{%
  \includegraphics{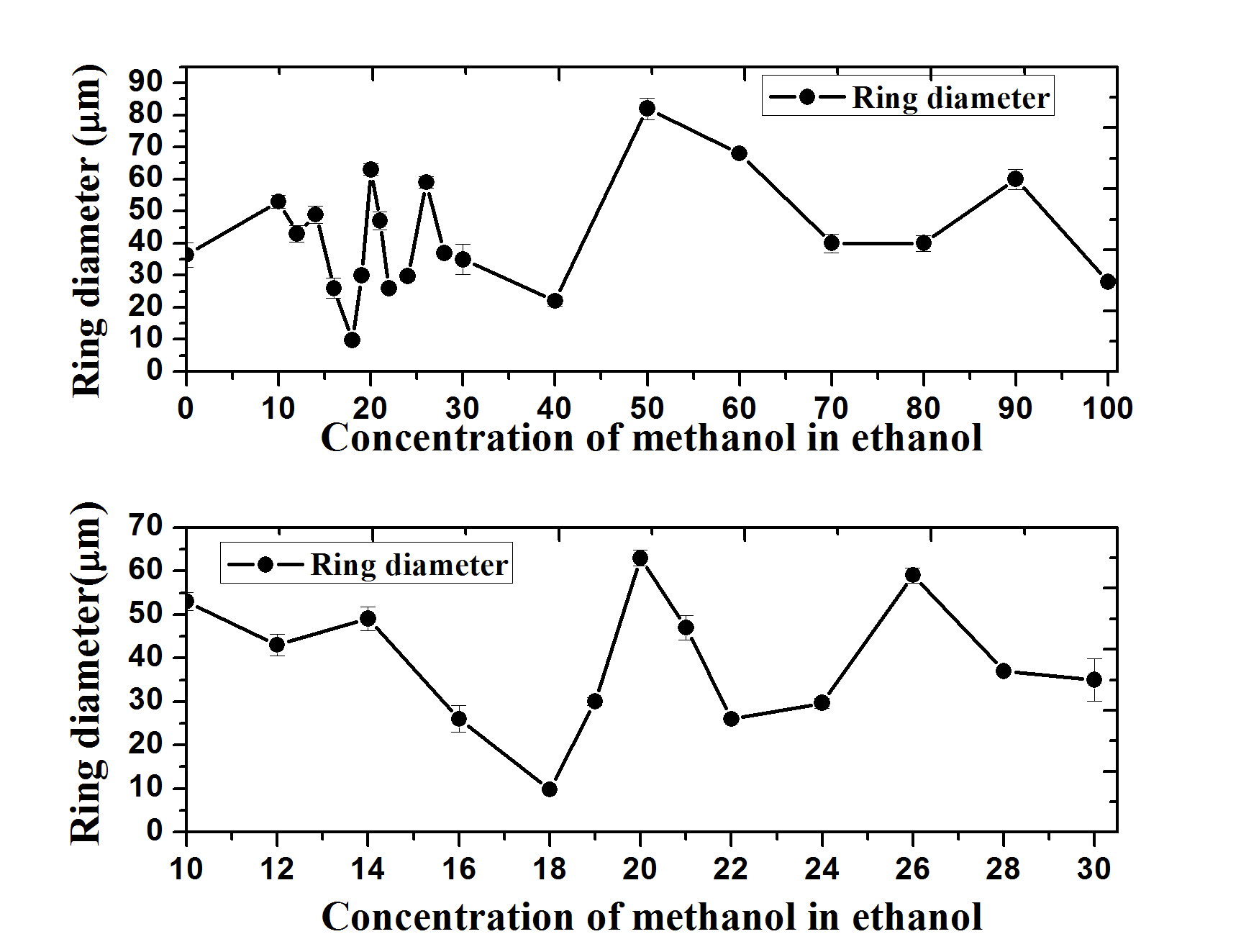}
	\vspace{0.5cm}
}
\caption{Variation of ring diameter of pores on constrained surfaces patterned under various concentration of methanol in ethanol.(Top: entire concentration region; bottom: concentration region $10\%$ to $30\%$ of methanol)}%
\label{8}
\end{figure}	
The dynamic viscosity of ethanol and methanol are $0.001095 Ns/m^{2}$ and $0.00056 Ns/m^{2}$ respectively. With low dynamic viscosity, methanol droplets shows a high the contact line velocity.  This results in the formation of more aligned hexagonal rings compared with ethanol of comparatively low dynamic viscosity. (Supporting 3D image of pure methanol on constrained surface is shown in Figure~\ref{5-a}.) Thus, highly ordered hexagonal ring patterns are obtained with saturated vapors of higher concentration of methanol.  This also explains the complex variation in pore ring diameter with varying concentration of methanol (plots are shown in Figure~\ref{8}). It should also be noted that when the horizontal component of the force caused by the Marangoni stress overcomes the Young's force, the Marangoni effect also has its influence on the depinning of contact line. \\ 
 
The energy barrier associated with the pinning and depinning of contact lines in the constrained surface is determined by the size of the grooves.  The competition between the Young unbalanced force and the anchoring forces of the grooves is thought to dominate the pinning/depinning of contact lines \cite{duursma}.  Since the groove size was maintained constant, the depinning/pinning of contact line is solely determined by the properties of condensing droplets. \\
\begin{figure}[h]
\resizebox{0.5\textwidth}{!}{%
  \includegraphics{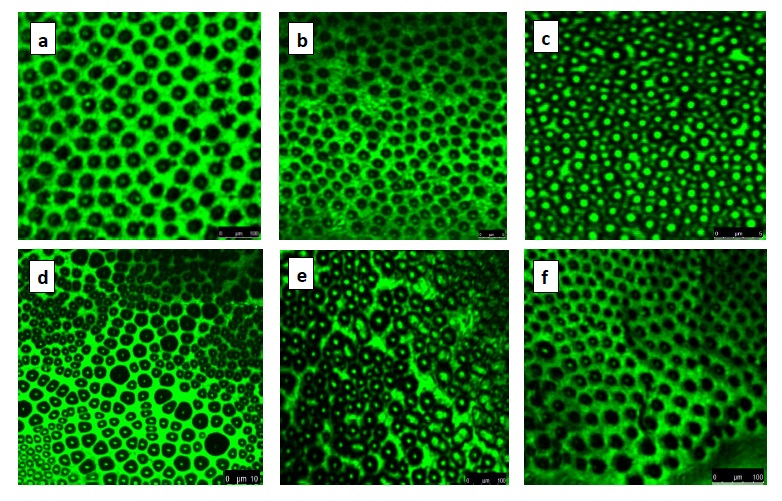}
  \vspace{0.5cm}
}
\caption{CLSM images of PDMS films over smooth substrate prepared in various concentration of binary vapors of ethanol-methanol. a)$12\%$ of methanol  b) $18\%$ of methanol c) $19\%$ of methanol d)$21\%$ of methanol e) $26\%$ of methanol f) $28\%$ of methanol.}%
\label{9}
\end{figure}
\begin{figure}[h]
\resizebox{0.5\textwidth}{!}{%
  \includegraphics{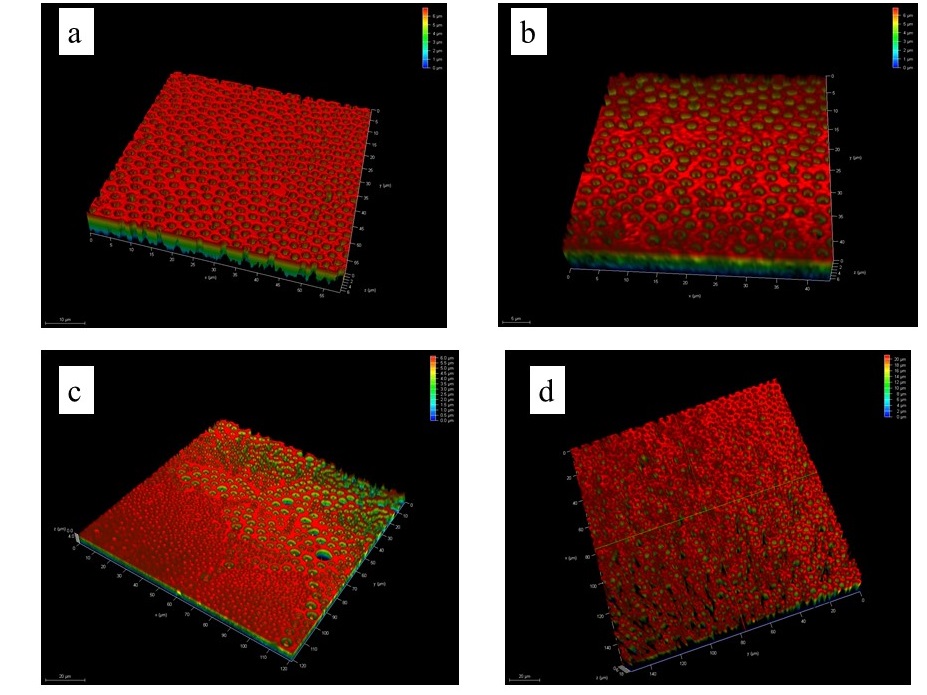}
	\vspace{0.5cm}
}
\caption{3D CLSM images of PDMS films over smooth substrate prepared in various concentration of binary vapors of ethanol-methanol. a)$12\%$ of methanol  b) $18\%$ of methanol c) $21\%$ of methanol d)$24\%$ of methanol.}%
\label{9-a}
\end{figure}

\begin{figure}[h]
\resizebox{0.5\textwidth}{!}{%
  \includegraphics{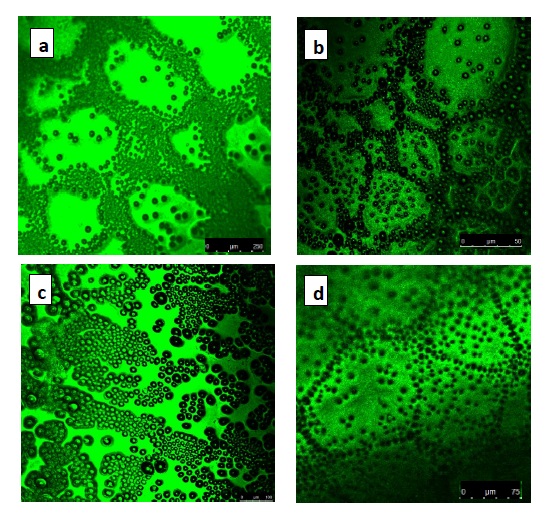}
	\vspace{0.5cm}
}
\caption{CLSM images of PDMS films over constrained substrate prepared in various concentration of binary vapors of ethanol-methanol. a)$12\%$ of methanol  b) $16\%$ of methanol c) $19\%$ of methanol d)$21\%$ of methanol.}%
\label{10}
\end{figure}
\begin{figure}[h]
\resizebox{0.5\textwidth}{!}{%
  \includegraphics{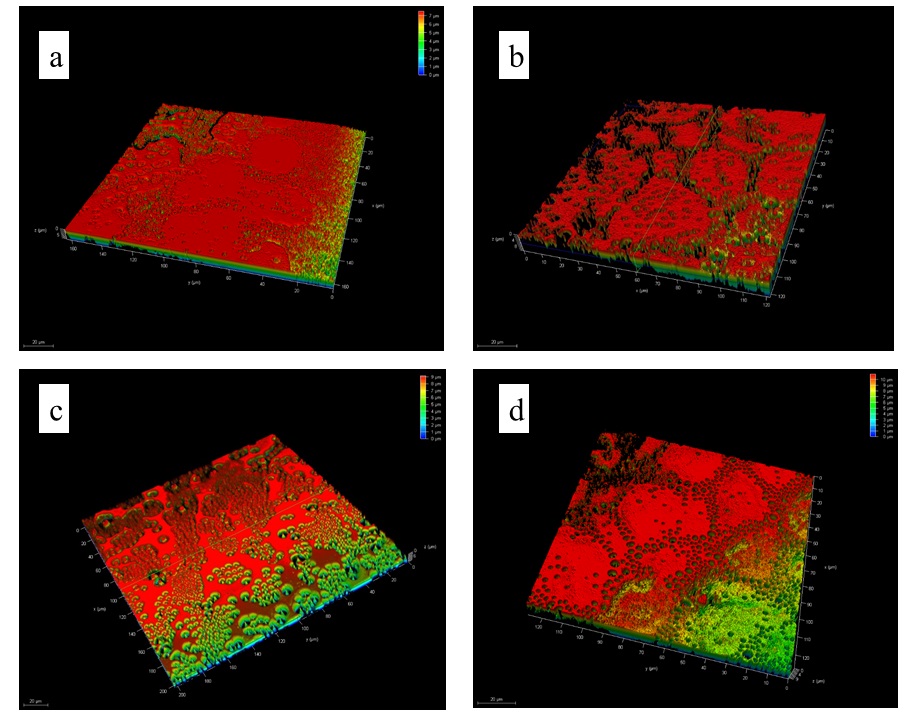}
	\vspace{0.5cm}
}
\caption{3D CLSM images of PDMS films over constrained substrate prepared in various concentration of binary vapors of ethanol-methanol. a)$14\%$ of methanol  b) $16\%$ of methanol c) $19\%$ of methanol d)$26\%$ of methanol.}%
\label{10-a}
\end{figure}
\subsection{Patterning under low concentration of methanol}
\label{subsec:3.4}
To gain a better understanding of the role of surface tension and other properties mentioned above, the patterning is performed under lower concentration of methanol in ethanol-methanol binary system, where complexity in their properties is reported \cite{madhu}. Figure~\ref{9} shows the polymer films over smooth substrate patterned under saturated vapors of $10\%$ to $30\%$ of methanol(in ethanol-methanol binary system)and Figure~\ref{9-a} shows the 3D CLSM images of the same.  It is evident that concentration increase of methanol beyond $10\%$ favors the formation of ordered hexagonal array of pores. It is observed that the pore distribution shows complex variation with varying concentration of methanol.  It is also observed from the figure that the pore size, pore depth and pore diameter varies significantly with varying surface tension of the saturated vapors. The maximum pore size and pore depth is observed for the films patterned under saturated vapor of $22\%$ of methanol. The pores are found to be highly deformed from circularity and at $19\%$, $21\%$ and $28\%$ of methanol, the circularity of the pores increases. This further proves that the surface tension of binary vapors plays a major role in tuning the shape, size and distribution of the pores.  \\

On examining the patterned constrained surface (Figure~\ref{10}), the films patterned under $16\%$, $21\%$ and $26\%$ of methanol show ordered hexagonal ring of pores.  It is evident from these results that not only the Young's unbalanced force but also the force due to constraints has its role in determining the alignment of pores. Figure~\ref{10-a} shows the 3D CLSM images of patterns on constrained surfaces.  In summary, it is observed that the pore diameter depends on the surface tension of ethanol-methanol vapors and pore patterns are influenced by the pinning/depinning of contact lines in addition commonly attribution to the properties of the solvent.  \\

\subsection{Pore aspect ratio}
\label{subsec:3.5}
\begin{figure}[h]
\resizebox{0.5\textwidth}{!}{%
  \includegraphics{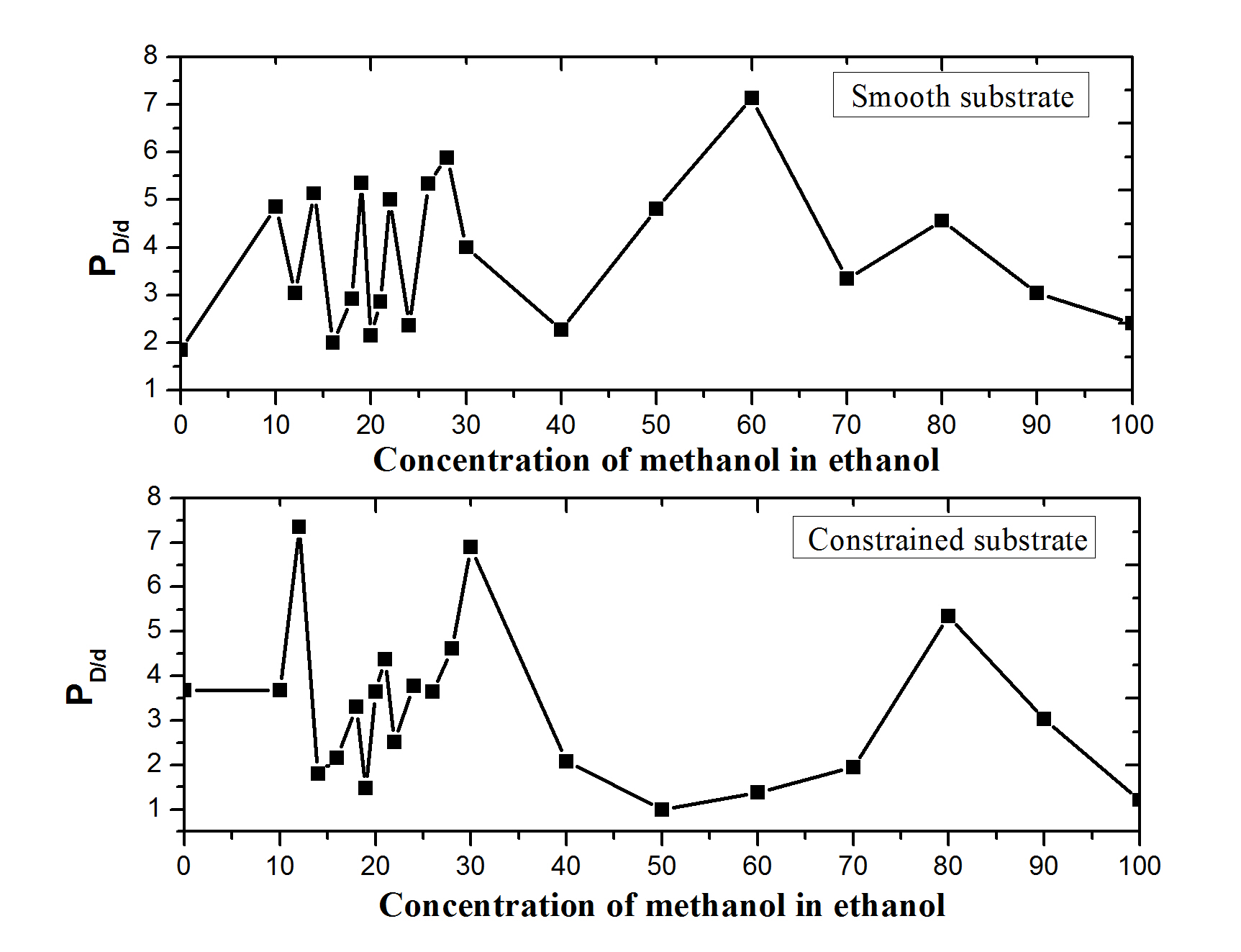}
	\vspace{0.5cm}
}
\caption{Variation of depth-diameter aspect ratio for smooth(top) and constrained(bottom) surfaces patterned with various concentration of methanol in ethanol. Uncertainty is less than the symbol size}%
\label{11}
\end{figure}

Pore aspect ratio $(P_{D/d})$ is defined as the ratio between the depth of the pores(D) and diameter(d) of the pores.i.e., 
\begin{equation}
P_{D/d} = D/d 
\label{eq:}
\end{equation}
 $P_{D/d}$ for all patterned surfaces is calculated and is shown in Figure~\ref{11}.  From the figure it is observed that the maximum depth-diameter aspect ratio obtained with smooth and constrained patterned surfaces are 7.1 and 7.3 respectively. For smooth surface, this maximum value is achieved under $60\%$ of methanol and for constrained surface it is achieved under $12\%$ of methanol.  The pore aspect ratio is observed to be strongly influenced the intermolecular interactions of binary vapors.  \\
\subsection{Contact angle studies}
\label{subsec:3.6}
\begin{figure}[h]
\resizebox{0.5\textwidth}{!}{%
  \includegraphics{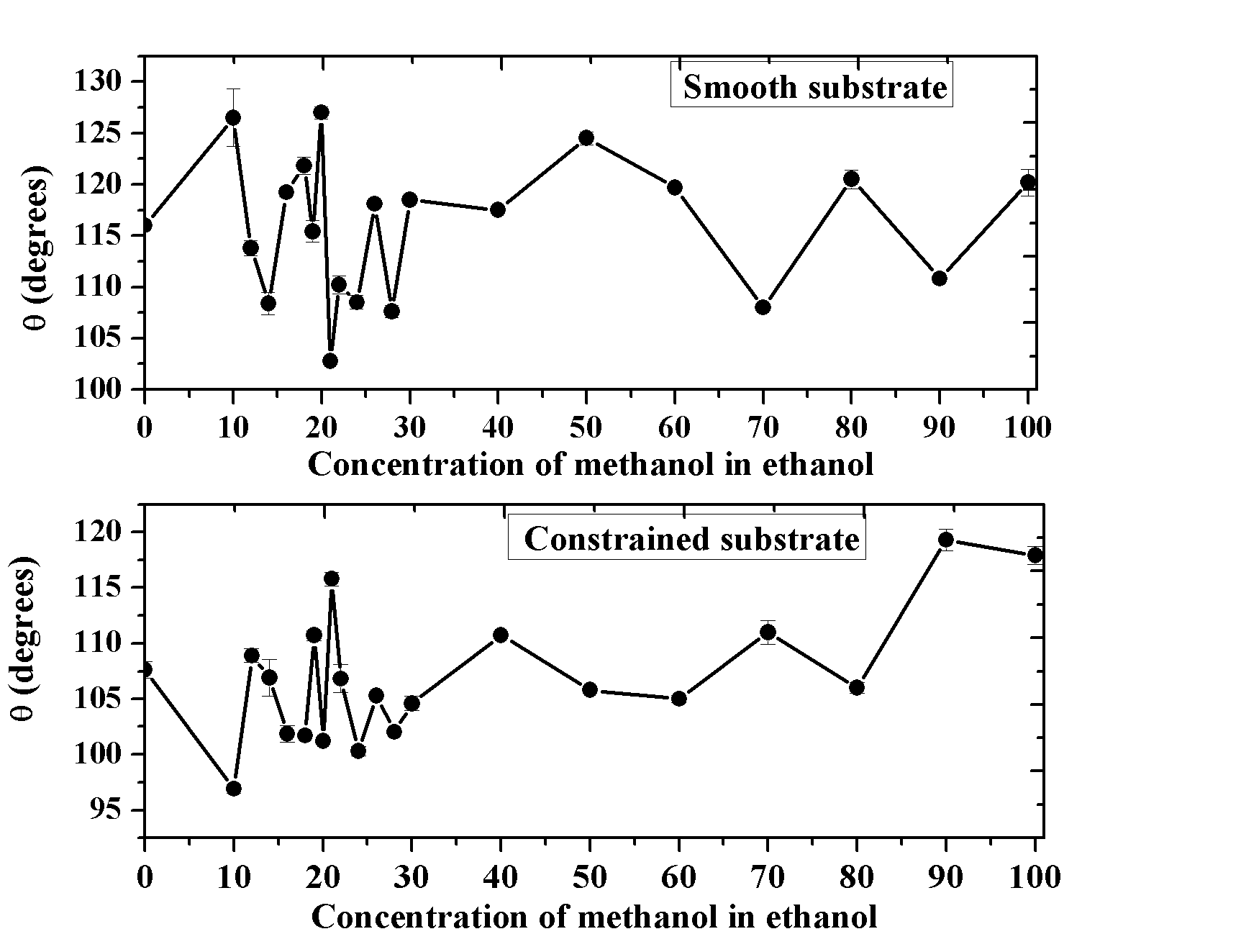}
	\vspace{0.5cm}
}
\caption{Variation of water contact angle for smooth(top) and constrained(bottom) surfaces patterned with various concentration of methanol in ethanol.}%
\label{12}
\end{figure}

Contact angle studies are performed for the smooth and constrained patterned surface using Goniometer with distilled water as the reference liquid. Surfaces exhibited hydrophobicity with water contact angle of more than $95^{o}$. A variation in contact angle is observed for films patterned under saturated vapors of binary system of ethanol-methanol as observed in Figure~\ref{12}. Maximum contact angle of $128^{o}$ is observed for smooth patterned surface under saturated vapor of $19\%$ of methanol.  For constrained surface, the maximum contact angle is observed at $22\%$ of methanol.  This variation of water contact angle confirms the variation in surface roughness which in turn attributed to the formation of different pore morphology on films patterned under various surface tension of vapor atmosphere. It is also observed from the results that the hydrophobicity/hydrophilicity can be achieved by the proper choice of the concentration of methanol.   \\
		
\section{Conclusion}
In conclusion, five main results are observed from this study. 
1.It has been demonstrated that it is possible to form ordered array of self-assembled droplets of methanol by the addition of small amount of ethanol without the use of any surfactants/water. This is attributed to the hydrogen bond between the ethanol and methanol and the ability of ethanol to bind strongly to water since it is more hydrophilic than methanol. 
2. The difference in pattern formation over smooth and constrained substrates indicates a strong influence by the depinning of the three phase contact line formed between substrate, binary liquid vapor and PDMS solution.  
3. Pore diameter is observed to be a function of surface tension of the ethanol-methanol binary concentration. 
4. Pore aspect ratio is influenced by intermolecular interactions of binary vapors. 
4. The complex variation in the shape, size and distribution of the pores are observed in the same concentration range where the complex anomalous variation in properties of ethanol-methanol binary system is reported earlier. 
5. The contact angle studies indicates that the hydrophobicity/hydrophilicity can be achieved by the proper choice of the concentration of methanol. And hence it can find its applications as self-cleaning surfaces, micro-fluidic devices, and water-repellant surfaces.

\section{Acknowledgement}

The authors thank Naval Research Board of India for the Contact Angle Goniometer.

%\section*{References} 

\section{Authors contributions}
All the authors were involved in the preparation of the manuscript.
All the authors have read and approved the final manuscript.

\end{document}